\documentclass{elsart3p}

\usepackage{graphicx}

\usepackage{amssymb}

\begin{document}

\begin{frontmatter}



  \title{Unified description of charge and spin excitations of stripes
  in cuprates.}


\author[a]{J. Lorenzana}
\author[b]{and G. Seibold}

\address[a]{SMC-INFM,ISC-CNR and Dipartimento di Fisica, Universit\`a di 
Roma ``La Sapienza'', Piazzale Aldo Moro 2, I-00185 Roma, Italy}
\address[b]{Institut f\"ur Physik, BTU Cottbus, PBox 101344,
         03013 Cottbus, Germany}

\begin{abstract}
We study stripes in cuprates within the one-band and 
the three-band Hubbard model. Magnetic and charge excitations
are described within the time-dependent Gutzwiller approximation.
A variety of experiments  (charge profile from resonant soft
X-ray scattering, incommensurability vs. doping,
optical excitations, magnetic excitations, etc.) are described within 
the same approach. 
 \end{abstract}

\end{frontmatter}

\section{Introduction} 

If the mechanism of high-$T_c$ superconductivity is electronic, 
to understand the excitation spectrum is as important as
to understand phonons was important to develop BCS theory. 
In this regard charge and spin inhomogeneous states, often found in 
strongly correlated systems, are 
interesting because they can support new collective modes, ``electronic
phonons'', that  would not be present in a weakly interacting
fluid.  For example stripes in cuprates have many oscillatory modes 
that can couple with carriers and eventually lead to pairing and
superconductivity or anomalous Fermi liquid properties.

Cuprates, unlike BCS superconductors,   
are unique and therefore their remarkable properties may well
be rooted on details. For example a recent study shows that
superconductivity in ${\rm La_{2-x}Sr_xCuO_4}$ appears only when 
incommensurate low energy scattering parallel to the CuO bond is present
and not, when the scattering is on the diagonals (at very low doping), 
or when the system is in a more conventional Fermi liquid phase
at high doping\cite{wak99wak04}, thus the excitation spectra must be
understood in detail. 
In the last years we have developed a method to compute collective
excitations of inhomogeneous strongly correlated systems based on a
time-dependent Gutzwiller approximation (GA), named 
GA+RPA\cite{sei01,lor03,lor05,sei05,sei06}.  

Many theoretical descriptions of cuprates are suited for only one
 experiment. 
Here we show that within the same approach one can reproduce a variety of 
experimental features of ${\rm La_{2-x}Sr_xCuO_4}$ in the charge and
magnetic channel. Results in the one-band Hubbard model (1BHM) 
 and the three-band Hubbard model (3BHM) 
are quite similar. We compare results from both models. 

For the 3BHM we use the following LDA
parameter set:  $\epsilon_{p}-\epsilon_{d}=3.3$ eV for
the splitting between the diagonal energies of a hole in the
copper $d$- and oxygen $p$ orbital, $t_{pd}=1.5$ eV 
($t_{pp}=0.6$ eV) for the $p$-$d$ ($p$-$p$) hopping
integral, $U_{d}=9.4$ eV ($U_{p}=4.7$ eV) for the
repulsion between two holes on the same Cu (O) orbital and
$U_{pd}=0.8$ eV for the Cu-O repulsion\cite{mcm90}. For the 1BHM
results reported here 
we use $t=353.7$meV for the nearest neighbor hopping,
$U=8t$ for the effective on-site repulsion, and 
$t'=-0.2t$ for the next-nearest neighbor hopping.

\section{Ground state properties}
\begin{figure}[t]
\includegraphics[width=7cm,clip=true]{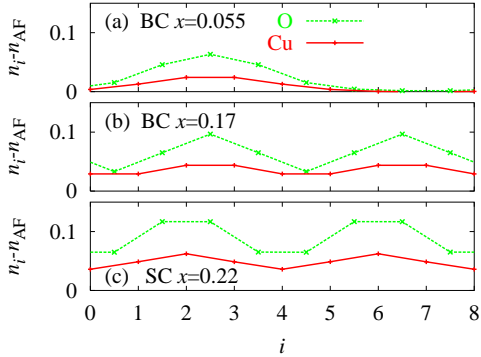}
\caption{Hole charge minus undoped charge in the 3BHM as a
 function of atomic position in the direction
  perpendicular to the stripes for various doping $x$, symmetry (BC
  or SC) and charge periodicity $d=9$ (a); and  $d=4$ (b) and (c). }
\label{fig:rdx}
\end{figure}

Static properties are computed within  the Gutzwiller approximation.  
Probes which are sensitive to the Cu
and O atomic character need to be addressed with the 3BHM.  In Fig.~\ref{fig:rdx} we show the
evolution of the charge profile with doping in the 3BHM for metallic
stripes parallel to the CuO bond. 
The charge profile in real space has a width of about 4 lattice 
sites\cite{lor02b}.
The width of the stripes defines two regimes: for low doping (a), 
the stripes do not overlap and therefore are weakly
interacting, for high doping (b),(c) stripes overlap. 

Recently Abbamonte and collaborators have shown that our charge
profile predicted in the 3BHM\cite{lor02b} is in
excellent agreement with the profile measured with resonant
soft X-ray scattering\cite{abb05}.

\begin{figure}[tb]
\includegraphics[width=8 cm]{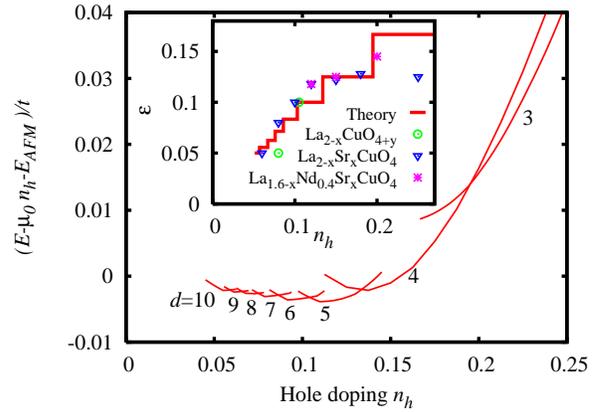}
\caption{Energy per site as a function of doping for stripes in the
  1BHM.  The curves are labeled
  by the charge periodicity $d$. 
For clarity we subtracted the energy of the AFM
  solution and the line $\mu_0 x$ with $\mu_0=-1.6t$.
Different  choices of $\mu_0$ correspond to different choices of the origin of
the  energy of the single particle states and do not change the
physics.   The inset
reports the incommensurability as obtained from the present calculation
(line) compared with experiment~\protect\cite{tra95}.}
\label{edxstr} 
\end{figure}

The overall shape of the  charge profile in the 1BHM is
similar to the one in the 3BHM (see  Ref.~\cite{sei04a}).
One also finds a non overlapping regime ($d>4$), and an overlapping  regime
($d\leq 4$). 

For large and negative $-t'/t$ (non shown here) one finds that checkerboard 
states are favored\cite{sei06c}. 

In Fig.~\ref{edxstr} we show the energy for metallic stripes as a
function of doping for the 1BHM. In each curve the charge
periodicity perpendicular to the stripe is fixed and takes integer
values (from left to right) $d=10-3$.  In this computations stripes
are bond-centered (BC). The energy for site-centered (SC) stripes is
slightly lower at small doping but the differences decreases with the
system size. At optimum doping both structures become degenerate.

A small difference with respect to the 3BHM is that in the
latter BC stripes are more stable at low doping and become
quasidegenerate with SC ones at optimum doping. Thus the role of SC
and BC is interchanged. We expect the three-band results to be more
reliable in this respect.

In the inset we plot the incommensurability vs. doping 
taking into account the range of stability of each solution and
compared with experiments from Refs.~\cite{tra95}.
The result is a Devil's staircase. 
 Up to $x\approx 1/8$ the plateaux are short and correspond to a 
number of added holes per unit length
 along the stripe close to $\nu\sim 0.5$. As doping increases one
 jumps from one solution to the other and the density of stripes increases
 with doping. This explains the behavior of the incommensurability
 $\epsilon=1/(2d)=x/(2\nu)\approx x$ as seen in neutron scattering 
experiments in this doping range. For $x>1/8$ the right branch of the $d=4$ 
 solution is more stable than
 the  $\nu\approx 0.5$ and $d=3$ solution due to the overlap effect
 and one gets a wider plateau explaining the saturation of the
 incommensurability seen in the experiment. 
 
Some details are slightly different in the 3BHM. The 
$\epsilon=1/8$ plateau starts at lower doping and ends at higher
 doping improving the agreement with experiment\cite{lor02b}.  
For doping $x>1/8$ holes populate the stripes and
therefore the baseline in Fig.~\ref{fig:rdx} increases.

 For doping $x>0.195$ ($x>0.225$)  in the one-band (three-band)  model
   we find the $d=3$ stripe ($\epsilon=1/6\approx 0.17$)
 to become the lowest energy solution. At this doping 
a variety of different solutions becomes  close in energy and the
   present saddle-point approximation brakes down.
Probably around this doping a quantum melting of stripes takes
place.


It is also possible that the $d=4$ stripe solution phase separates
with the overdoped Fermi liquid skipping the $d=3$ solution. This
scenario will also produce a large $\epsilon=1/8$ plateau in the 
incommensurate scattering. In this case long range Coulomb effects
have to be taken into account\cite{lor02}.

One can see from the plot of the energy in Fig.~\ref{edxstr} 
that the electron chemical potential
$\mu=-\partial E/\partial x$ is approximately constant in the low
doping regime (weakly interacting stripes) 
whereas it decreases in the overlapping stripe regime. 
The same behavior was found in a dynamical mean-field study of the
one-band Hubbard model\cite{fle00}.
This is in
qualitative agreement with experiment\cite{ino97}  although we find a larger
rate of change of $\mu$ with doping than in the experiment. 
It is also possible that this is due to phase separation between 
$d=4$ stripes and the overdoped Fermi liquid plus long-range Coulomb effects\cite{lor02}.

\section{Magnetic excitations}

\begin{figure}[tbp]
$$\includegraphics[width=6 cm,clip=true]{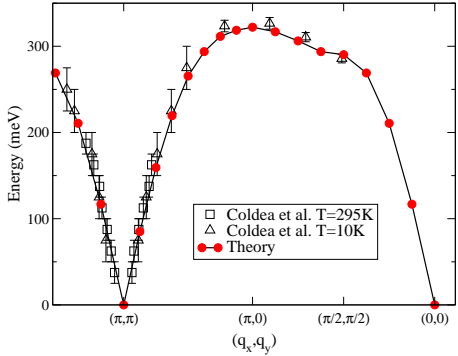}$$
\caption{Dispersion relation of the low-energy transverse
excitations.  We show the experimental result for La$_2$CuO$_4$
after Ref.~\protect\cite{col01} and the  GA+RPA result.}
\label{coldea} 
\end{figure}

The dispersion relation for
magnetic excitations in the insulator computed within GA+RPA applied to
the 1BHM are shown
in Fig.~\ref{coldea} and is in excellent agreement with the
experiment. The dispersion is weakly sensitive to $t'/t$ and it is
strongly sensitive to $U/t$. 
A larger $U/t$ produces a flatter dispersion
in the $(\pi,0)$ - $(\pi/2,\pi/2)$ direction which does not agree
with the experiment\cite{lor05}. 

The collective excitations in the stripe phase with the same parameter
set have been reported in Ref.~\cite{sei05}. One obtains 
a resonance energy at 65 meV to be compared with the experimental
value  55 meV\cite{tra04}. A slight decrease of $U$ improves the
agreement of the magnetic excitations with experiment as reported in
Ref.~\cite{sei06} but produces a
charge transfer gap in the insulator which is 10\% too low. The
present parameter set is a compromise between the charge and magnetic
channel.
Details of the magnetic response and the doping dependence are discussed
in  an accompanying paper\cite{sei06dresden}.



\begin{figure}[tbp]
\includegraphics[width=7cm,clip=true]{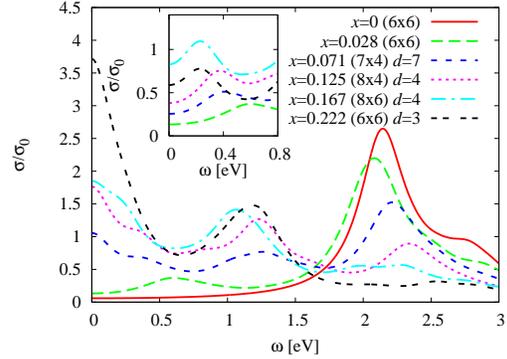}
\caption{Optical conductivity labeled by doping, system size and, in
  the case of stripes, interstripe distance.
 The units of conductivity are given by $\sigma_0=3.6
\times 10^2 (\Omega {\rm cm})^{-1}$ with a background
dielectric constant $\epsilon_b=2.6$
 The curve labeled
$x=0.028$ corresponds to the single-hole solution. For
larger dopings the figure is an average over the electric field directions
parallel and perpendicular to the BC stripes. The inset shows the MIR band excluding the Drude component. We used a Lorentzian broadening
 of 0.2eV.}
\label{fig:sdw}
\end{figure}

\section{Optical conductivity}

The optical conductivity is qualitatively similar in 
the three-band  and  1BHM. In Fig.~\ref{fig:sdw} we
show the result in the 3BHM.  The charge transfer (CT) gap
in the insulator is close to 2eV in good agreement with the experiment. 
In the 1BHM the CT gap is also at  2eV. 
This is mainly determined by our value of
$U=2.8eV$ confirming our parameter choice.

Doping induces a doping dependent mid-infrared (MIR) 
band and Drude weight at zero energy since our stripes are metallic.
The  MIR band is due to collective lateral fluctuations of the stripe.
These lateral fluctuations form a band of excitations as for a violin
string and the MIR band corresponds to the zero momentum component.   

The ``string'' fluctuation band is massive due to pinning of the commensurate stripes to the 
underlaying lattice. However, both the experiment\cite{uch91} and our 
computations\cite{lor03}
show that the mass decreases with doping.  This is rooted in  the
quasi-degeneracy found among BC and SC stripes and indicates that
stripes form a floating phase at optimum doping. The MIR band produces
an absorption at low energy which explains the failure of the
Drude model at optimum doping. This indicates that  
anomalous Fermi liquid properties are rooted in the low energy
excitations of the floating phase. 
\begin{table}[tbp]
  \caption{Selected excitations in meV in the one-band Hubbard model within the
    GA+RPA and in Experiment. Values marked with $*$ are extrapolated.   \label{tab:expthe}
  }
\begin{tabular}{ccc|ccc}   
                      &   Momentum & $x$ & Theory       &Exp.   & Ref.\\
\hline
CT gap                &     0     &  0     & $\sim 2000$    &$\sim 2000$  & \cite{uch91}   \\
MIR band              &     0     &  0.08   & $\sim  500$    &$\sim 250^*$   & \cite{uch91} \\ 
Magnon          & $(\pi/2,\pi/2)$ &  0     & 293                 & $286\pm 5 $ & \cite{col01}   \\
Magnon          & $(\pi,0)$       &  0     & $326$                 & $333\pm 7^*$  & \cite{col01}  \\
Resonance       &     0           &  1/8   & 65           &  55   & \cite{tra04} \\
  \end{tabular}
\end{table}

In the 1BHM the MIR band is at energy $\sim 0.5$ eV for doping
0.08 which is higher than in the experiment (Table~\ref{tab:expthe}).  
Experimentally the
band is located at $\sim 0.5$ eV at very  low doping and softens with
doping faster than what we found.
This may be due to a finite size effect since
our optical conductivity computations are done in system sizes much
smaller than the computations of magnetic properties 
($16\times 4$ and  $40\times40$ respectively.)

Table~\ref{tab:expthe} summarizes the energy of selected excitations
in the 1BHM with the present parameter set. It is remarkable
that such a simple model with only three parameters is able to provide
a reasonable description of a variety of excitations in different
channels plus several ground state properties.

\section{Discussion and Conclusions}

In cuprates true magnetic long-range order is sometimes
detected in experiment\cite{tra95} however, more often, Bragg peaks are not
found. Two possibilities arise: the system may be close to a
quantum critical point in the disordered phase (dynamic stripes) or the 
system may be in the ordered side of the transition but long-range
order is not observed because of disorder (glassy stripes). 
Dynamic stripes have quasi long-range order in time and space whereas 
glassy stripes at low temperatures have practically long-range order in time
but short range order in space as in a structural glass. 
This latter situation is clear for example in the magnetic channel at
low doping, in muon spin-relaxation experiments where below a certain temperature
the system develops a static (on the muon time scale) magnetic 
field which is not necessarily accompanied by Bragg peaks\cite{ker03},
thus the spin rotational symmetry is broken without long range order. 
In these quantum or classically disordered cases our results apply as 
long as the energy scale is not too low (typically  tenths of meV). 
How this  reflects on the Fermi surface is discussed
separately\cite{dic06dresden}.

To conclude we have shown that the GA+RPA approximation allows for a unified description of
collective modes in the charge and magnetic channel of striped cuprates. Results are similar in
the one-band and the 3BHM although some details may
differ in which case the 3BHM is expected to be more
accurate. These modes are likely to play an important role in the
anomalous properties of cuprates. 



\end{document}